\documentclass[a4paper]{article}
\usepackage{sita2023}
\usepackage{amsmath}
\usepackage{bm}
\usepackage{amssymb}
\usepackage{autobreak}
\usepackage{url}
\usepackage{here}
\usepackage{multirow}
\usepackage[table,xcdraw]{xcolor}

\usepackage[utf8]{inputenc}
\usepackage{newunicodechar,graphicx}
\DeclareRobustCommand{\okina}{%
  \raisebox{\dimexpr\fontcharht\font`A-\height}{%
    \scalebox{0.8}{`}%
  }%
}

\DeclareMathOperator{\VT}{VT}
\DeclareMathOperator{\con}{con}

\title{
The Tight Upper Bound for the Size of Single Deletion Error Correcting Codes in Dimension 11
}

\author{
Kazuhisa Nakasho
    \thanks{
        Graduate School of Science and Technology for Innovation,\\
        Yamaguchi University,
        2-16-1, Tokiwa-dai, Ube, Japan\\
        E-mail: {\tt nakasho@\allowbreak
            yamaguchi-u.\allowbreak
            ac.\allowbreak
            jp}
    }
    \and
Manabu Hagiwara
    \thanks{
        Graduate School of Science,
        Chiba University,\\
        1-33, Yayoicho, Inage-ku, Chiba, Japan\\
        E-mail: {\tt hagiwara@\allowbreak
          math.\allowbreak
          s.\allowbreak
          chiba-u.\allowbreak
          ac.\allowbreak
          jp}
    }
    \and
Austin Anderson
    \thanks{
        Math and Sciences,
        Kapi\okina olani Community College,\\
        4303 Diamond Head Rd, Honolulu, United States\\
        E-mail: {\tt austina@\allowbreak
          hawaii.\allowbreak
          edu}
    }
    \and
J. B. Nation
    \thanks{
        Department of Mathematics,
        University of Hawaii,\\
        2500 Campus Road, Honolulu, United States\\
        E-mail: {\tt jb@\allowbreak
          math.\allowbreak
          hawaii.\allowbreak
          edu}
    }
}

\abstract{
A single deletion error correcting code (SDECC) is a set of fixed-length sequences consisting of two types of symbols, 0 and 1, such that the original sequence can be recovered for at most one deletion error. The upper bound for the size of SDECC is expected to be equal to the size of Varshamov-Tenengolts (VT) code, and this conjecture had been shown to be true when the code length is ten or less. In this paper, we discuss a method for calculating this upper bound by providing an integer linear programming solver with several linear constraints. As a new result, we obtained that the tight upper bound for the size of a single deletion error correcting code in dimension 11 is 172.
}


\begin{document}
\maketitle

\section{Introduction}

A deletion error occurs when some symbols from words are missing, leading to shorter words. We focus on codes made up of binary sequences, specifically 0s and 1s. There is a class of codes, known as deletion error correcting codes (DECCs), that can restore the original word even after certain deletion errors. If a code can handle up to $n$ missing symbols, it is termed an $n$-deletion error correcting code ($n$-DECC). For a single missing symbol, it is referred to as a single deletion error correcting code (SDECC).


Consider the following code of length 5:
$$
\{00000, 10001, 01010, 11011, 11100, 00111 \} .
$$
For each word in this code, the series resulting from a single deletion error can be listed as follows:
\begin{itemize}
    \item $00000 \to 0000$
    \item $10001 \to 0001, 1001, 1000$
    \item $01010 \to 1010, 0010, 0110, 0100, 0101$
    \item $11011 \to 1011, 1111, 1101$
    \item $11100 \to 1100, 1110$
    \item $00111 \to 0111, 0011$
\end{itemize}
Observing these series of length 4, we find that there are no duplicates. This means that even if a single symbol is deleted from this code, it can still be uniquely restored to its original word of length 5. In essence, this is an SDECC of length 5.


One of the typical examples of SDECC is the VT codes.
A VT code of length $n$ has the words $\bm{x} = x_1 x_2 \cdots x_n$, where $a$ is an integer and
$$
\VT_a(n) :=
\{
\bm{x} \mid x_1 + 2 x_2 + \cdots + n x_n = a \pmod{n + 1}
\}.
$$
The aforementioned code of length 5 is equal to $\VT_0(5)$.
Some salient properties of the VT codes will be discussed in the next section.


Deletion errors result in the loss of positional information as well as the deleted symbols. This complexity has led to many open problems about DECCs, and even in the simplest case of SDECCs, several interesting open problems remain. One notable open problem is about the maximum size of an SDECC. It is believed that the maximum size of an SDECC of length $n$ matches the size of $\VT_0(n)$. This conjecture had been confirmed for values up to $n \le 10$. Furthermore, Albert No~\cite{no2019nonasymptotic} in 2019, by using mixed integer linear programming (MILP), proved the upper bound for $n=11$ is $|\VT_0(11)| + 1\ (= 173)$ or less.


In this paper, we detail our contributions as follows: Firstly, we introduced several constraints that an SDECC must meet, framed as linear integer inequalities. Secondly, by applying an integer linear programming (ILP) solver to these constraints, we confirmed the conjecture that the maximum size of an SDECC of length $n=11$ is equal to $|\VT_0(11)|\ (=172)$. Additionally, we modified the set of constraints provided to the ILP solver to assess the effectiveness of each constraint in addressing the problem.

\section{Preliminaries}
\subsection{Terminologies}

Let $m,n$ be positive integers in this subsection.
{
\setlength{\leftmargini}{15pt}
\begin{itemize}
\item $M(n)$ : The maximum size of SDECC of length $n$.
\item $\bm{x}, \bm{y}, \bm{z},\dots$ : Binary sequences made up of 0s and 1s. In particular, to emphasize that the length is $n$, we use the notation $\bm{x}^{(n)}$ or $\bm{x} \in \{0,1\}^n$. When representing $\bm{x}$ as a sequence of symbols, we write it as $\bm{x} = x_1 x_2 \dots x_n$ with subscripts ordered from 1 upwards.
\item $\bm{0}, \bm{1}$ : Sequences made up solely of 0s and 1s, respectively. To specify a sequence of length $n$, we use the notation $\bm{0}^{(n)}$ or $\bm{0} \in \{0,1\}^n$.
\item $d_L(\bm{x},\bm{y})$: Levenshtein distance between $\bm{x}$ and $\bm{y}$ represents the minimum total number of insertions and deletions required to transform $\bm{x}$ into $\bm{y}$.
\item $W_H(\bm{x})$: The Hamming weight of $\bm{x}$, i.e., the number of 1s in the sequence $\bm{x}$.
\item $r(\bm{x}, b)$: The count of $b$-runs within $\bm{x}$, where $b \in \{0,1\}$, refers to sequences consisting solely of repeated $b$ values. For instance, in $\bm{x} = 1010001001$, there are three 0-runs: 0, 000, and 00. Thus, $r(\bm{x}, 0) = 3$.
\item $dS_t(\bm{x})$: A deletion surface with radius $t$ centered around $\bm{x}$. This can be expressed as:
$$
dS_t(\bm{x}) := \{\bm{y} \in \{0,1\}^{n-t} \mid d_L(\bm{x},\bm{y}) = t\} .
$$
For $t=1$, we simply denote $dS_1$ as $dS$.
\item $V(\bm{x},C)$: A 0-1 integer variable that indicates if the sequence $\bm{x}$ is a part of the set $C$. It is 1 if $\bm{x} \in C$ and 0 if $\bm{x} \not\in C$. Using this notation, the number of elements in the set $C$ can be expressed as:
$$
|C| = \sum_{\bm{x} \in \{0,1\}^n} V(\bm{x},C) .
$$
\item $\con(\bm{x}_1, \dots, \bm{x}_k)$ : The concatenation of the series $\bm{x}_1, \dots, \bm{x}_n$.
For example,
$$
\con(\bm{x}^{(m)}, \bm{y}^{(n)}) = x_1 x_2 \cdots x_m y_1 y_2 \cdots y_n .
$$
\end{itemize}
}

\subsection{Properties of VT codes}

VT codes are notable not just for their tolerance to single deletions, but also for a unique feature known as perfectness. An SDECC of length $n$ is called perfect if, when single deletion errors are applied to each word, the resulting set of sequences collectively covers the entire set of sequences of length $n-1$. This means it satisfies the following two conditions:
$$
\bigcup_{\bm{x} \in \VT_a(n)} dS(\bm{x}) = \{0,1\}^{n-1} ,
$$
$$
\forall \bm{x}, \bm{y} \in \VT_a(n),\ \bm{x} \ne \bm{y} \to dS(\bm{x}) \cap dS(\bm{y}) = \emptyset .
$$
For instance, when considering all elements of $\VT_0(5)$, their deletion surfaces precisely partition the entire set of sequences of length 4 into 6 distinct subsets.


The following properties are known about the size of VT codes:
\begin{align*}
|\VT_0(n)| = \frac{1}{2(n+1)} \sum_{\substack{d \mid n+1 \\ 2 \nmid d}} \phi(d) 2^{(n+1)/d} ,
\end{align*}
\begin{align*}
|\VT_1(n)| = \frac{1}{2(n+1)} \sum_{\substack{d \mid n+1 \\ 2 \nmid d}} \mu(d) 2^{(n+1)/d} ,
\end{align*}
where $\phi$ and $\mu$ are Euler's totient function and Möbius function, respectively.
Moreover, for any $a \in \mathbb{Z}$, the following property is satisfied:
\begin{align*}
|\VT_0(n)| \ge |\VT_a(n)| \ge |\VT_1(n)| .
\end{align*}
Table \ref{vt_size} shows the values of $|\VT_a(n)|$ for small code length $n$.
\begin{table}[H]
\caption{$|\VT_a(n)|$}
\label{vt_size}
\begin{tabular}{c|ccccccccc}
\hline
$n$\textbackslash $a$ & 0  & 1  & 2  & 3  & 4  & 5  & 6  & 7  & 8  \\ \hline
1 & 1  & 1  &    &    &    &    &    &    &    \\
2 & 2  & 1  & 1  &    &    &    &    &    &    \\
3 & 2  & 2  & 2  & 2  &    &    &    &    &    \\
4 & 4  & 3  & 3  & 3  & 3  &    &    &    &    \\
5 & 6  & 5  & 5  & 6  & 5  & 5  &    &    &    \\
6 & 10 & 9  & 9  & 9  & 9  & 9  & 9  &    &    \\
7 & 16 & 16 & 16 & 16 & 16 & 16 & 16 & 16 &    \\
8 & 30 & 28 & 28 & 29 & 28 & 28 & 29 & 28 & 28 \\
\hline
\end{tabular}
\end{table}

\section{Preceding Studies}

Classic papers of Levenshtein~\cite{Levenshtein1966, Levenshtein2002} and Sloane~\cite{sloane2000single} remain insightful references on the maximum size problem of SDECC.
A prominent result concerning the maximum size of SDECC is encapsulated in the inequality:
$$
\frac{2^n}{n+1} \leq |\VT_0(n)| \leq M(n)  \leq \frac{2^n-2}{n-2} .
$$
This is detailed in Levenshtein~\cite{Levenshtein1966} and Kulkarni and Kiyavash~\cite{Kulkarni2013}.
The lower bound is given by VT codes, which possess the remarkable ability to divide the $2^n$ words of length $n$ into $n+1$ equivalence classes, each of roughly equal size.

Table \ref{table:maxsdecc} lists some known values, using improved upper bounds from
\cite{Kulkarni2013}, see also \cite{Cullina2014}.
Sloane confirmed the optimality of $|\VT_0(n)|$ for $n \leq 7$, and David Applegate did so for $n=8$~\cite{sloane2000single}.
The cases for $n=9$ and $n=10$ appear to be addressed in Butenko et~al.~\cite{Butenko2009}.
Albert No reduced the upper bounds for $n=11$ and $n=12$ to 173 and 320, respectively~\cite{no2019nonasymptotic}.

\begin{table}[H]
\centering
\caption{The maximum size of SDECC}
\label{table:maxsdecc}
$\begin{array}{c c c c }
\hline
 n   & |\VT_0(n)|   & M(n) & \text{known upper bound}\\
\hline
\rule{0ex}{3ex}
 2     &2  &2  &2    \\
 3     &2  &2  &2    \\
 4     &4  &4  &4    \\
 5     &6  &6  &6    \\
 6     &10  &10  &10    \\
 7     &16  &16  &16    \\
 8     &30 &30  &30  \\
 9     &52 &52  &52  \\
 10   &94 &94   &94 \\
 11   &172 &172   &172 \\
 12   &316 &?   &320 \\
 13   &586 &?  &593  \\
 14   &1096   & ? &1104 \\
 15   &2048    & ? &2184 \\
\hline
\end{array}$
\end{table}


\section{Methods}\label{sec:methods}
\setcounter{subsection}{-1}

Our approach transforms the problem of determining the upper bound size of SDECC into ILPs by introducing linear constraints that the SDECC must meet. ILP aims to find a set of non-negative integers $(x_1, \dots, x_n)$ that maximizes the linear expression
\begin{align*}
\sum_{i=1}^n c_i x_i
\end{align*}
given $m$ pairs of linear inequalities 
\begin{align*}
\sum_{i=1}^n a_{ij} x_i \le b_j \quad (j = 1,\dots, m)
\end{align*}
as constraints. Consequently, the problem of finding the maximum size of an SDECC can be framed as locating an SDECC $C$ such that 
\begin{align*}
\sum_{\bm{x} \in \{0,1\}^n} V(\bm{x},C)
\end{align*} 
is at its peak. We will now detail the linear constraints used for the ILPs. In subsequent discussions, $V(x,C)$ corresponding to the maximum-sized SDECC of length $n$ will be simply referred to as $V_x$.

\subsection{Constraint 0}

This fundamental constraint ensures that words in an SDECC can be recovered after single deletion errors. Specifically, it mandates that the deletion surfaces of any two distinct words do not overlap:
\begin{align*}
dS(x) \cap dS(x') = \emptyset .
\end{align*}
This requirement can be recast into linear inequalities:
\begin{align*}
\forall \bm{y} \in \{0,1\}^{n-1}, \sum_{\bm{y} \in dS(\bm{x})} V_{\bm{x}} \le 1 .
\end{align*}
While the above inequality is a necessary condition for $V_x$ to ensure $C$ is an SDECC, relying solely on this condition for the ILP solver can be inefficient. In subsequent subsections, we will explore constraints that, while logically redundant, can enhance the ILP solver's performance.

\subsection{Constraint 1}

This constraint is straightforward: it asserts that the maximum size of the SDECC should be at least as large as the size of the VT code:
\begin{align*}
\sum_{\bm{x} \in \{0,1\}^n} V_{\bm{x}} \ge \VT_0(n) .
\end{align*}
As we will discuss later, this constraint played a pivotal role in enhancing performance during our experiments.

\subsection{Constraint 2}

Let $C$ be an SDECC with $\bm{x} \in C$. If there is a $\bm{y} \in C$ such that $\bm{y} \ne \bm{x}$ and $dS(\bm{y}) \subset dS(\bm{x})$, then the set $C' = C \setminus \{\bm{x}\} \cup \{\bm{y}\}$ is also an SDECC, as it meets Constraint 0. Given that $dS(\bm{x})\cap dS(\bm{y})\ne \emptyset$, $\bm{x}$ and $\bm{y}$ cannot both be in the SDECC. Thus, we can infer that a largest SDECC includes $\bm{y}$, which has the smaller deletion surface. This relationship can be formulated as the following linear inequalities:
\begin{align*}
\forall \bm{x}, \bm{y} \in \{0,1\}^n,\ \bm{x} \ne \bm{y},\ dS(\bm{y}) \subset dS(\bm{x}) \to V_{\bm{x}} = 0 .
\end{align*}

\subsection{Constraint 3}

For any sequence $\bm{x}$, if its Hamming weight $W_H(\bm{x})$ is 1, then the condition $dS(\bm{0}) \subset dS(\bm{x})$ holds. On the other hand, if $dS(\bm{0}) \cap dS(\bm{x}) \ne \emptyset$, then $W_H(\bm{x}) = 1$. Based on this, we can deduce that $\bm{0}$ is always a part of a largest SDECC. The same applies to $\bm{1}$. These observations can be represented as the following linear inequalities:
$$
V_{\bm{0}} = V_{\bm{1}} = 1 .
$$

\subsection{Constraint 4}

For any SDECC $C$, the bit-flipped code defined as $C' = \{\bm{1} - c \mid c \in C\}$ is also an SDECC. This implies that for any integer $i$ in the range $0 \le i \le n$:
$$
\{c \in C \mid W_H(c) = i\} = \{c' \in C' \mid W_H(c') = n-i\} .
$$
From this, in terms of maximum size, we can deduce:
$$
\sum_{0 \le i \le n/2} |\alpha_i| \ge \sum_{n/2 \le i \le n} |\alpha_i|
$$
where $\alpha_i := \{c \in C \mid W_H(c) = i\}$. This relationship can be formulated as the following linear inequality:
$$
\sum_{\bm{x} : 0 \le W_H(\bm{x}) \le n/2} V_{\bm{x}} \geq \sum_{\bm{x} : n/2 \le W_H(\bm{x}) \le n} V_{\bm{x}} .
$$

\subsection{Constraint 5}
For integers $0 \le w, \alpha, \beta \le n$, let us define a set comprising bit sequences of length $n$:
$$
\mathcal{W}_n(w,\alpha,\beta) := \{\bm{x} \mid W_H(\bm{x}) = w, r(\bm{x},0) = \alpha, r(\bm{x},1) = \beta\} .
$$
Put simply, $\mathcal{W}_n(w,\alpha,\beta)$ encompasses all bit sequences of length $n$ that have a Hamming weight of $w$, $\alpha$ 0-runs, and $\beta$ 1-runs.

Consider a sequence $\bm{x} \in \mathcal{W}_n(w,\alpha,\beta)$. What would be the bit sequences that belong to $dS(\bm{x})$? If we remove 0 from $\bm{x}$, the Hamming weight remains the same, but $\alpha$ new bit sequences emerge. Conversely, if we delete 1 from $\bm{x}$, the Hamming weight decreases by one, resulting in $\beta$ bit sequences. From this observation, we can derive the following linear inequality, taking into account a set with Hamming weight $w$ and sequence length $n-1$:
\begin{align*}
&\sum_{0 \le \alpha_0, \beta_0 \le n} \alpha_0 (\sum_{\bm{x} \in \mathcal{W}_n(w, \alpha_0, \beta_0)} V_{\bm{x}})\\ + &\sum_{0 \le \alpha_1, \beta_ 1 \le n} \beta_1 (\sum_{\bm{x} \in \mathcal{W}_n(w+1, \alpha_1, \beta_1)} V_{\bm{x}}) \le \binom{n-1}{w}.
\end{align*}
It is worth noting that $\mathcal{W}_n(w,\alpha, \beta) = 0$ when the absolute difference $|\alpha - \beta|$ exceeds 1, which stems from the inherent structure of the runs in the sequence. Additionally, if $\beta > w$, then $\mathcal{W}_n(w,\alpha, \beta) = 0$. This is a consequence of the characteristics of 1-run and Hamming weights.

Let us consider the case when $w=1$. In this scenario, the sum spans both $\mathcal{W}_n(1, \alpha_0, \beta_0)$ and $\mathcal{W}_n(2, \alpha_1, \beta_1)$. Given Constraint 3, we can deduce that $\mathcal{W}_n(1,\alpha_0, \beta_0) = \emptyset$. As a result, the subsequent linear inequality emerges:
$$
\sum_{1 \le \alpha_1 \le 3, 1 \le \beta_1 \le 2} (\beta_1 \sum_{\bm{x} \in \mathcal{W}_n(2,\alpha_1, \beta_1)} V_{\bm{x}}) \le n-1 .
$$

\subsection{Constraint 6}

Let $p, q, r$ be non-negative integers such that $p + q + r = n$ and $p + q > 0$. Now, consider bit strings $\bm{u} \in \{0,1\}^p$ and $\bm{v} \in \{0,1\}^q$. We define 
$$
D(C, \bm{u}, \bm{v}) := C \cap \{ \con(\bm{u}, \bm{x}, \bm{v}) \mid \bm{x} \in \{0,1\}^r \}.
$$
Given that $C$ is an SDECC, for any distinct $\bm{x}, \bm{y} \in D(C, \bm{u}, \bm{v})$, their deletion surfaces do not overlap: $dS(\bm{x}) \cap dS(\bm{y}) = \emptyset$. Therefore, for any $\bm{u} \in \{0,1\}^p$ and $\bm{v} \in \{0,1\}^q$, the inequality $|D(C, \bm{u}, \bm{v})| \le M(r)$ holds true. This can be expressed as the following linear inequalities:
$$
\forall \bm{u} \in \{0,1\}^p, \forall \bm{v} \in \{0,1\}^q, 
\sum_{\bm{x} \in \{0,1\}^r} V_{\con(\bm{u}, \bm{x}, \bm{v})} \le M(r).
$$

\section{Experiments}

We incorporated the constraints outlined in section \ref{sec:methods} into an ILP solver to evaluate its performance. The hardware and software specifications used for the experiment are detailed below:
\begin{itemize}
    \item CPU: AMD Ryzen 9 5900X 12-Core Processor (3.7GHz)
    \item Memory: 128GB
    \item Software: Gurobi Optimizer 10.0.0
\end{itemize}
The source code is available at the following GitHub repository:
\begin{itemize}
    \item \url{https://github.com/aabaa/deletion_code}
\end{itemize}

\subsection{Length of Code n=10}

Table \ref{table:n10} presents the experimental outcomes for $n=10$. Within this table, constraints provided to the solver are indicated with $\checkmark$ mark. Additionally, the time required to solve the problem is displayed. The ``Ratio" represents the relative time, with the shortest solver time normalized to 1. Since Constraint 0 was consistently applied in all scenarios, it is not explicitly listed in Table \ref{table:n10}.

\begin{table}[H]
\caption{Constraints and solver performance (n=10)}
\label{table:n10}
\centering
\begin{tabular}{rccccccccrr}
\hline
                     & \multicolumn{6}{c}{Constraints}   &                                                      &                         \\ \cline{2-7}
\multirow{-2}{*}{No.} & 1 & 2 & 3 & 4 & 5 & 6 & \multirow{-2}{*}{Time[s]}                               & \multirow{-2}{*}{Ratio} \\ \hline
1)  &   &   &   &   &   &   & 1602.0                                               & 12.07 \\ \hline
2)  & $\checkmark$ &   &   &   &   &   & 289.9                                                & 2.18  \\ \hline
3)  & $\checkmark$& $\checkmark$&   &   &   &   & 291.9                                                & 2.20  \\ \hline
5)  & $\checkmark$&   & $\checkmark$&   &   &   & 291.2                                                & 2.19  \\ \hline
6)  & $\checkmark$&   &   & $\checkmark$&   &   & 134.7 & 1.01  \\ \hline
7)  & $\checkmark$&   &   &   & $\checkmark$&   & 146.5 & 1.10  \\ \hline
8)  & $\checkmark$&   &   &   &   & $\checkmark$& 303.9                                                & 2.29  \\ \hline
9)  & $\checkmark$&   &   & $\checkmark$& $\checkmark$&   & 133.5 & 1.01  \\ \hline
10) & $\checkmark$&   &   & $\checkmark$&   & $\checkmark$& 271.5                                                & 2.05  \\ \hline
11) & $\checkmark$&   &   &   & $\checkmark$& $\checkmark$& 266.8                                                & 2.01  \\ \hline
12) & $\checkmark$&   &   & $\checkmark$& $\checkmark$& $\checkmark$& 310.2                                                & 2.34  \\ \hline
13) & $\checkmark$& $\checkmark$& $\checkmark$&   &   &   & 292.1                                                & 2.20  \\ \hline
14) & $\checkmark$& $\checkmark$& $\checkmark$& $\checkmark$&   &   & \textbf{132.7}                & \textbf{1.00}  \\ \hline
15) & $\checkmark$& $\checkmark$& $\checkmark$&   & $\checkmark$&   & 146.5                         & 1.10  \\ \hline
16) & $\checkmark$& $\checkmark$& $\checkmark$&   &   & $\checkmark$& 165.3                                                & 1.25  \\ \hline
17) & $\checkmark$& $\checkmark$& $\checkmark$& $\checkmark$& $\checkmark$&   & 134.0                         & 1.01  \\ \hline
18) & $\checkmark$& $\checkmark$& $\checkmark$& $\checkmark$&   & $\checkmark$& 150.5                                                & 1.13  \\ \hline
19) & $\checkmark$& $\checkmark$& $\checkmark$&   & $\checkmark$& $\checkmark$& 190.3                                                & 1.43  \\ \hline
20) & $\checkmark$& $\checkmark$& $\checkmark$& $\checkmark$& $\checkmark$& $\checkmark$& 150.0                                                & 1.13  \\ \hline
\end{tabular}
\end{table}


From result 1), we observe that by simply applying Constraint 0, a solution for $n=10$ can be achieved in a reasonably practical timeframe using a state-of-the-art ILP solver. However, result 2) demonstrates that by incorporating Constraint 1, which specifies the lower bound of the maximum size of SDECC, computational time can be substantially reduced. This suggests that this single condition can significantly constrict the search space.


Furthermore, it is evident that the impact of each constraint on computational time is not independent. As a result, it is challenging to expect the exact influence of each constraint on the problem. Yet, as displayed in Table \ref{table:n10}, for $n=10$, the combination in case 14) appears nearly optimal.


Interestingly, case 20), where all constraints were applied, did not produce the quickest results, contrary to what one might expect. This suggests that overloading the solver with too many constraints can actually hinder its performance.

\subsection{Length of Code n=11}


\begin{table}[htbp]
\caption{Constraints and solver performance (n=11)}
\label{table:n11}
\centering
\begin{tabular}{rccccccccrr}
\hline
\multicolumn{1}{c}{\multirow{2}{*}{No.}} & \multicolumn{6}{c}{Constraints}   & \multirow{2}{*}{Time{[}s{]}} & \multirow{2}{*}{Ratio} \\ \cline{2-7}
\multicolumn{1}{c}{}                    & 1 & 2 & 3 & 4 & 5 & 6 &                              &                        \\ \hline
1) & $\checkmark$ &    &  &   &   &   & $\gg$ 1M & $\gg$ 27 \\ \hline
2) & $\checkmark$ &    &  & $\checkmark$&   &   & $\gg$ 1M & $\gg$ 27 \\ \hline
3) & $\checkmark$ &    &  &   & $\checkmark$&   & \textbf{36,181}    & \textbf{1.00}       \\ \hline
4) & $\checkmark$ &    &  & $\checkmark$& $\checkmark$&   & 55,152    & 1.52       \\ \hline
5) & $\checkmark$ & $\checkmark$ & $\checkmark$& $\checkmark$&   &   & $\gg$ 1M & $\gg$ 27 \\ \hline
6) & $\checkmark$ & $\checkmark$ & $\checkmark$& $\checkmark$& $\checkmark$&   & 55,074   & 1.52       \\ \hline
7) & $\checkmark$ & $\checkmark$ & $\checkmark$&   &   & $\checkmark$& 315,932  & 8.73       \\ \hline
8) & $\checkmark$ & $\checkmark$ & $\checkmark$& $\checkmark$& $\checkmark$& $\checkmark$& $\gg$ 1M & $\gg$ 27 \\ \hline
\end{tabular}
\end{table}


The experimental outcomes for $n=11$, as depicted in Table \ref{table:n11}, deviated from our expectations based on the $n=10$ results. Notably, Constraint 5 proved to be particularly effective for $n=11$. In contrast, Constraint 2 and 3 did not appear as effective, especially when compared alongside pairs like cases 2) and 5), and cases 4) and 6). Constraint 4 also did not seem to perform well in the context of cases 2), 3), and 4). The decline in performance due to the amalgamation of numerous constraints was more evident than for $n=10$. Cases 7) and 8) suggest that Constraint 6 might actually be detrimental to performance. It is believed that Constraint 6 becomes less favorable for the solver as the number of conditionals substantially grows with an increase in $n$.

\section{Discussion and Conclusion}

Our primary contributions include proposing several constraints for an ILP solver, which facilitated the determination of the maximum size of SDECC for $n=11$. However, regarding the upper bound size for $n=12$, we were unable to achieve a result lower than 320, as indicated by \cite{no2019nonasymptotic}.


To determine the maximum size of SDECC for $n=12$ using this method, we recognize the importance of identifying more effective constraints. As demonstrated by the experimental results in the preceding section, predicting the optimal combination of constraints for an ILP solver is challenging. The ideal combination largely hinges on the characteristics of the ILP solver in use. Different outcomes might arise if a solver other than the Gurobi Optimizer, which we employed in this research, is used. For optimization, it could be beneficial to select constraints based on a deep understanding of the solver's properties or even consider modifications to the solver algorithm itself.


In this study, we employed an ILP solver to establish the maximum size of SDECC, demonstrating that no solution satisfies the constraints. However, there is a risk in solely relying on ILP solvers with complicated implementation to justify the non-existence of solutions. To address this, a future direction could involve obtaining more reliable results using executable code produced by interactive theorem provers, as applied in \cite{Kondo2020formalization}.


\bibliographystyle{junsrt}
\bibliography{reference}
\end{document}